      [1999/12/01 v1.4c Il Nuovo Cimento]
\documentclass{cimento}

\usepackage[english]{babel}
\usepackage{epsfig}
\usepackage{subfigure}
\usepackage{graphicx}
\usepackage[table]{xcolor}

\def\babar{\mbox{\slshape B\kern-0.1em{\smaller A}\kern-0.1em
    B\kern-0.1em{\smaller A\kern-0.2em R}}}

\title{Study of Collins Asymmetries at BaBar}
\author{I.~Garzia\from{ins:x}}

\instlist{\inst{ins:x} INFN and University of Ferrara - Italy, for the \babar\ Collaboration.}

\PACSes{\PACSit{13.66.Bc}{Hadron production in $e^-e^+$ interactions}
\PACSit{13.87.Fh}{Fragmentation into hadrons}}
\begin{document}

\maketitle

\begin{abstract}
Transversity distribution describes the quark transverse polarization inside a transversely polarized nucleon. It is the less known  leading-twist piece of the QCD description of the partonic structure of the nucleon.
Transversity can be extracted from semi-inclusive deep inelastic scattering (SIDIS) where, however, it couples to a new, unknown fragmentation function, called {\mdseries\ttfamily Collins function}.
We present the preliminary results of the measurement of the azimuthal 
asymmetries in the process 
$e^+e^-\rightarrow q\overline{q}\rightarrow \pi \pi X$,
where the two pions are produced in opposite hemispheres.
These preliminary results are based on a data sample of
about 45 fb$^{-1}$, collected 
by the \babar\ experiment at a center-of-mass energy of 10.54 GeV, and
are compared with the Belle measurements.
\end{abstract}

\section{Motivation for extraction of Collins function in $e^+e^- $ annihilation}

\begin{figure}[!htb]
\centering
\subfigure[][ RF12] {
 \includegraphics[width=0.47\textwidth] {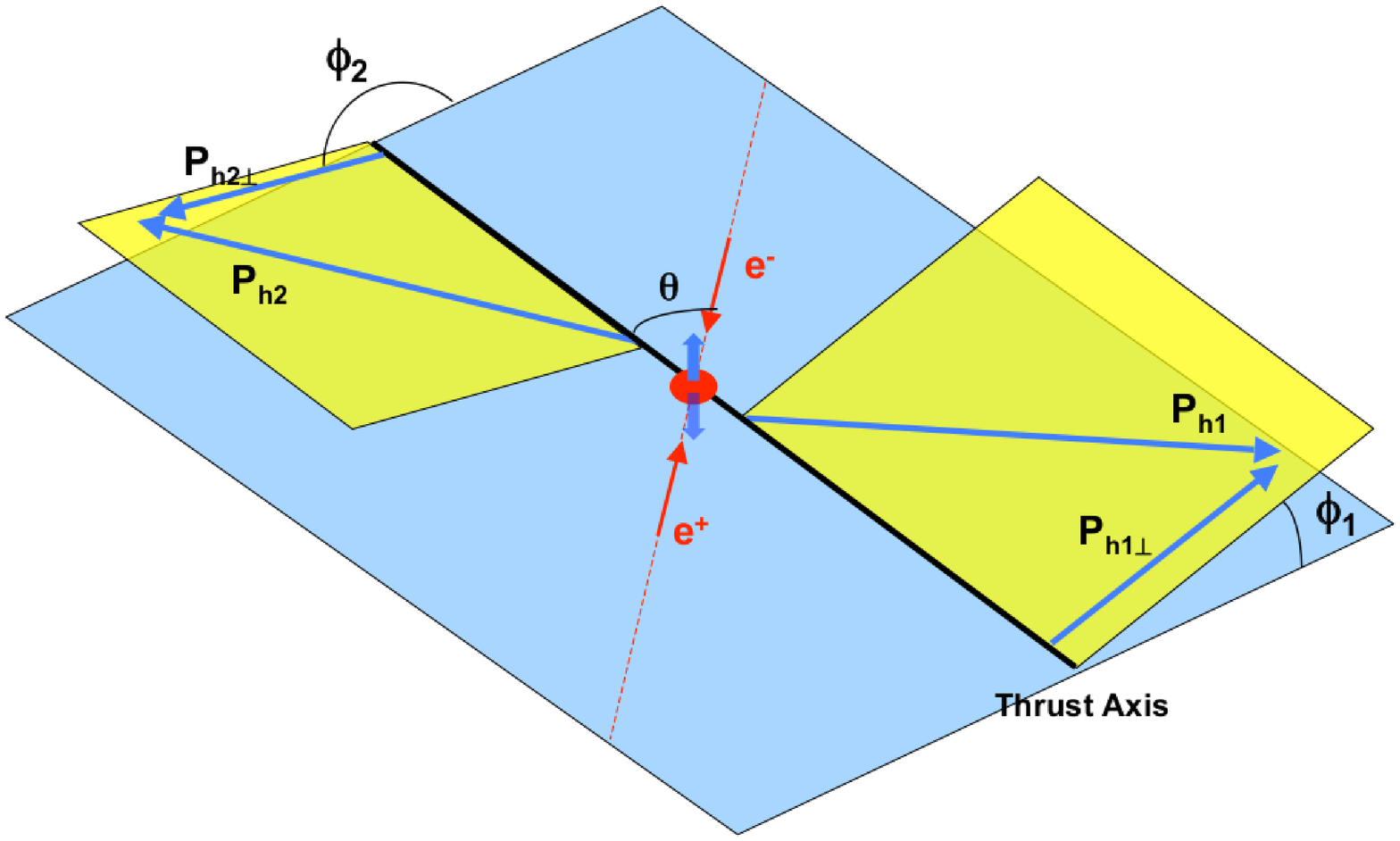} \label{fig:sdr12}
 }
\subfigure[][ RF0] {
   \includegraphics[width=0.47\textwidth] {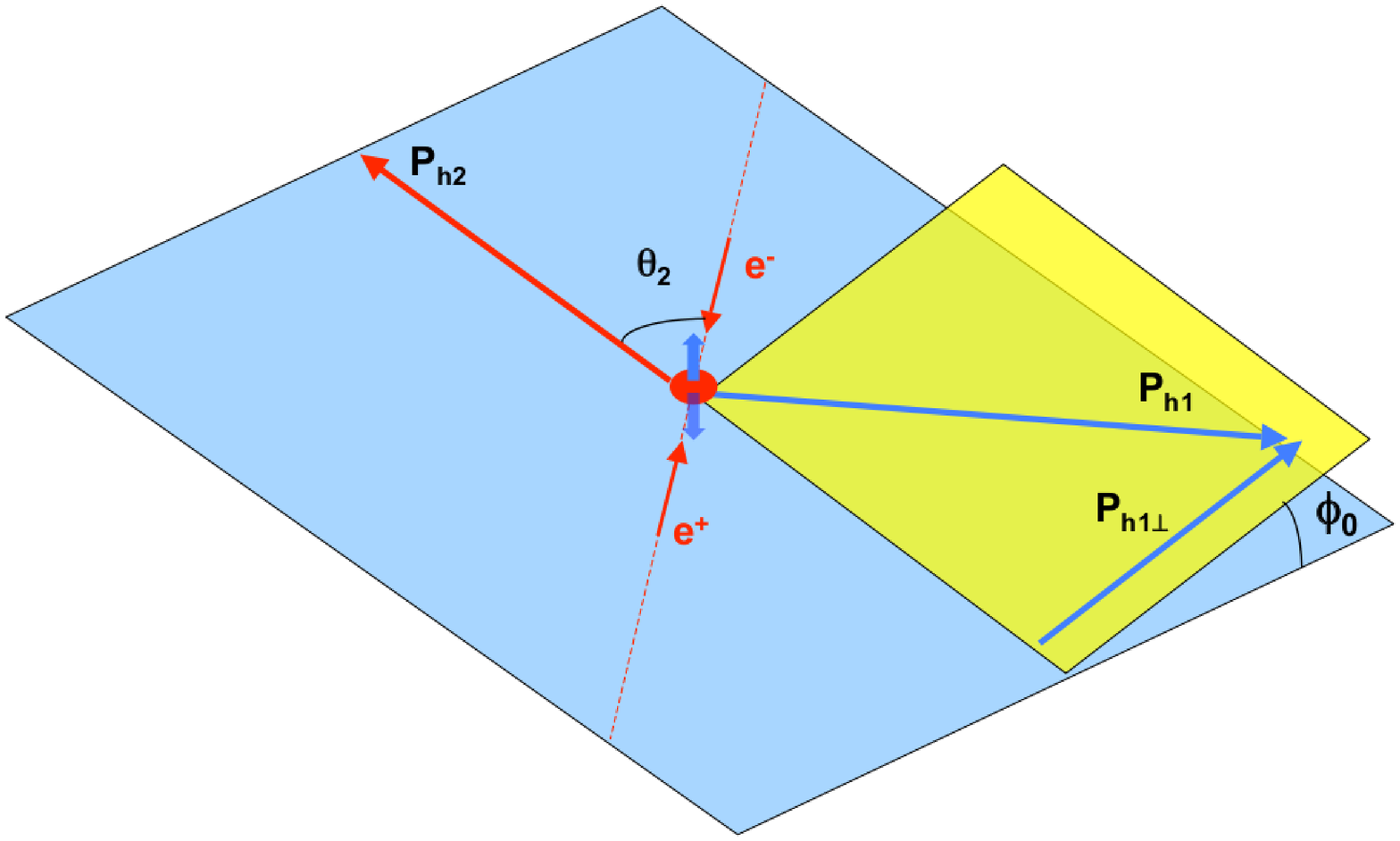} \label{fig:sdr0}
   }
\caption{(Color online) RF12 (a): $\theta=\theta_T$ is the angle between the beam and thrust axis; $\phi_{1,2}$ are the azimuthal angles between the scattering plane and the transverse momentum $P_{h\perp}$ around the thrust axis. 
RF0 (b): $\theta_2$ is the angle between the beam axis and the second hadron momentum; $\phi_0$ is the azimuthal angle between the plane spanned by the beam axis and the second hadron momentum $P_2$, and by the first hadron's transverse momentum $P_{1\perp}$.
All  tracks are boosted in the $e^+e^-$ center of mass frame.}
\label{fig:sdr}
\end{figure}

Transversity ($h_1$) is the less known function\cite{ref:ans} among the three parton distribution functions  needed for a complete description of the momentum and spin distribution of the quarks inside the nucleon.
We can measure $h_1$ in semi inclusive lepton-nucleon scattering
(Semi Inclusive Deep Inelastic Scattering or SIDIS),
in which the proton target is transversely polarized.
Using the factorization theorem, the SIDIS cross section is
\begin{eqnarray}
\sigma^{ep \rightarrow ehX} = \sum_{q} DF \times \sigma(eq\rightarrow eq) \times FF,
\end{eqnarray}
with DF the parton Distribution Function, and FF  the Fragmentation Function.
For transversely polarized target, DF is the transversity
distribution ($h_1$), and FF is the Collins fragmentation function ($H_1^{\perp}$).
Therefore, to extract $h_1$ we need to know $H_1^{\perp}$.
The measurement of the Collins function can be done in $e^+e^-$ annihilation
taking into account the process 
$e^+e^-\rightarrow q\overline{q}\rightarrow h_1 h_2 X$,
in which the two hadrons are detected in opposite jets.
A non zero value of the Collins FF produces an asymmetry
in the azimuthal distribution of the two hadrons around the 
fragmenting $q\overline{q}$ direction. 
The Collins asymmetry can be measured in two 
different reference frames.
In the first frame, called {\itshape thrust reference frame} or RF12, 
the azimuthal angles  $\phi_1$ and $\phi_2$ 
shown in fig.\ref{fig:sdr12} are calculated with respect to 
the thrust axis, that is the axis that maximize the longitudinal
momentum of the event.
In this frame the cross section is~\cite{ref:boer}:
\begin{eqnarray}\label{eq:rf12}
\frac{d\sigma(e^+e^-\rightarrow h_1h_2X)}{dz_1dz_2 d\cos\theta\, d\phi_1d\phi_2} &=& \sum_{q,\overline{q}}\frac{3\alpha^2}{Q^2}\frac{e_q^2}{4}z_1^2z_2^2 [(1+cos^2\theta)D_1^{(0)}(z_1) \overline{D}_1^{(0)}(z_2)+\\ \nonumber 
&+& sin^2(\theta)cos(\phi_1 + \phi_2) H_1^{\perp , (1)}(z_1) \overline{H}_1^{\perp , (1)}(z_2)]\\ \nonumber
\end{eqnarray}
In the second reference frame, 
{\itshape second hadron momentum frame} or RF0,
the azimuthal angle $\phi_0$ is calculated with respect to the second
hadron momentum, as shown in fig.\ref{fig:sdr0}, and
 the differential cross section is~\cite{ref:boer}:
\begin{eqnarray}\label{eq:rf0}
\frac{d\sigma (e^+e^+\rightarrow h_1 h_2 X)}{dz_1 dz_2 d^2 \mathbf{q}_T d\, \cos\theta_2 \,d\phi_0} &=& \frac{3\alpha^2}{Q^2}\; z_1^2 z_2^2 \; \left\{ A(y) \; \mathcal{F} (D_1 (z_1)  \overline{D}_1(z_2)) + \right. \nonumber \\
&+&  \left.  B(y) \; cos(2\phi_0) \mathcal{F} \left[ (2\hat{\mathbf{h}} \cdot  \hat{\mathbf{k}}_T \; \hat{\mathbf{h}} \cdot \hat{\mathbf{p}}_T -\hat{\mathbf{k}}_T \cdot \hat{\mathbf{p}}_T )  \frac{H_1^\perp (z_1)  \overline{H}_1^\perp (z_2)}{M_1M_2}   \right] \right\} 
\end{eqnarray}
where $A(y)=1/4(1+\cos^2\theta_2)$ and $B(y)=1/4(\sin^2\theta_2)$ in the $e^+e^-$ center of mass frame.

\noindent In eqs.~(\ref{eq:rf12}) and (\ref{eq:rf0}),
 $D_1$ is the unpolarized FF, $H_1^{\perp}$ is the Collins FF, $z_{1,2}$ are the fractional energy of the hadrons ($z=2E/Q$), $Q^2$ is the center-of-mass energy, and the angles are defined in fig.\ref{fig:sdr}. In conclusion, the Collins Asymmetry in $e^+e^-$ annihilation is proportional to $H_1^{\perp}(z_1)\times \overline{H}_1^{\perp}(z_2)$ and, therefore, we can obtain a independent measurement of this FF.

\section{Analysis strategy}
The preliminary measurement of Collins Asymmetries is performed
using a sample of data collected by the \babar\ experiment at the energy
of 10.54 GeV, that is 40 MeV below the nominal energy of the collider,
which corresponds to the peak of $\Upsilon(4S)$ resonance.
For this reason, we refer to this data sample as off-peak sample.

Assuming the thrust axis as $q\overline{q}$ direction and selecting pions in opposite hemispheres with respect to the thrust axis, we measure the azimuthal angles $\phi_1$, $\phi_2$, and $\phi_0$. 
In order to select the two jets topology, we require a thrust higher than 0.8.
In addition, we select pions coming from the primary vertex with a fractional energy 
higher than 0.2.
The total visible energy of the event (sum of the energies
of reconstructed charged and neutral particles) is required to be higher than 7 GeV.
The Collins asymmetries are obtained
by measuring the $\cos(\phi)$ modulation of the 
normalized azimuthal distributions of the selected pion pairs,
where $\phi=\phi_1+\phi_2$ or $\phi=2\phi_0$ 
for the two reference frames.
The asymmetries resulting by these distributions are largely affected by detector 
acceptance effects, making their measurement unreliable. 
We therefore perform suitable double ratios  of the asymmetries in order to
eliminate the detector effects and the first order of radiative corrections~\cite{ref:boer}.
In particular we use the ratio of the normalized distributions of 
 unlike sign ($R_{UL}$) over those of like sign ($R_L$) pion pairs.
 In addition, the possibility to select pion pairs with same charge or opposite
charge, allows to be sensitive to favored and disfavored fragmentation
functions.  
Favored FFs describe the fragmentation of a quark of flavor $q$ into a
hadron with a valence quark of the same flavor: i.e. $u\rightarrow
\pi^+$ and $d\rightarrow \pi^-$.   
Instead, we refer to  $u\rightarrow \pi^-$ and $d\rightarrow \pi^+$  as disfavored fragmentation processes.
Consider, for example, the production of unlike-sign charged pions
from a $u\overline{u}$ pair  : $e^+e^- \rightarrow u\overline{u} \rightarrow
\pi^{\pm} \pi^{\mp} X$. 
The pion pair can be either created through two favored fragmentation
processes or through two disfavored fragmentation processes.

\noindent Taking into account what said above, the double ratio 
can be written as
 
\begin{equation}\label{eqn:dr}
\frac{R^{UL}}{R^L} \, = \,
\frac{1+\frac{\sin^2\theta}{1+\cos^2\theta}
  \cos(\phi)\,G^{UL}}{1+\frac{\sin^2\theta}{1+\cos^2\theta}
  \cos(\phi)\,G^{L}}  
\, \simeq \, 1+ \frac{\sin^2\theta}{1+\cos^2\theta} \cos(\phi)
\left\{ G^{UL}-G^{L} \right\}\,, 
\label{eq:DR}
\end{equation}

\begin{eqnarray}
G^{UL}=\frac{\sum_q e^2_q \mathcal{F}(H_1^{fav} H_2^{fav}+ H_1^{dis}
  H_2^{dis}) }{\sum_q e^2_q (D_1^{fav}D_2^{fav} +
  D_1^{dis}D_2^{dis})},\,\,\,\,\,  
G^{L}=\frac{\sum_q e^2_q \mathcal{F}(H_1^{fav} H_2^{dis}+ H_1^{dis}
  H_2^{fav} )}{\sum_q e^2_q (D_1^{fav}D_2^{dis} +
  D_1^{dis}D_2^{fav})}\, ; \nonumber
\end{eqnarray}
 where $\theta$ is the polar angle of the thrust axis in the RF12 frame
 ($\theta=\theta_T$)
 or the polar angle of the second hadron momentum in the RF0 frame
 ($\theta=\theta_2$),
 as shown in fig.\ref{fig:sdr}.
 
Fitting the double ratio with a cosine function
\begin{eqnarray}
\frac{R_{UL}}{R_{L}}= \frac{N^{UL}(\phi)/<N^{UL}>}{N^{L}(\phi)/<N^{L}>}=P_0+P_1\cdot cos(\phi),
\end{eqnarray}
the $P_1$ parameter  contains only the Collins effect and higher order of radiative effects,
since acceptance and radiative contributions do not depend 
on the charge combination of the pion pairs.

\section{Study of systematic effects}
A crucial point for the measurement of Collins asymmetry is the identification of all the effects that can influence the azimuthal distribution of the pion pairs.
We study the influence of the particle identification, the possible dependence of the detector response on the pion charge, the presence of residual polarization of the beams, and other minor effects.

\subsection*{Asymmetry dilution due to the charm and $\tau$ decays} 
 Measured asymmetries are diluted by the presence of background sources 
like charm and $\tau$ decays. For this reason 
 we study and evaluate the correction
 to the asymmetries in order to obtain the true Collins effect.

\begin{figure}[!htb]
  \begin{center}
   \includegraphics[width=0.45\textwidth]{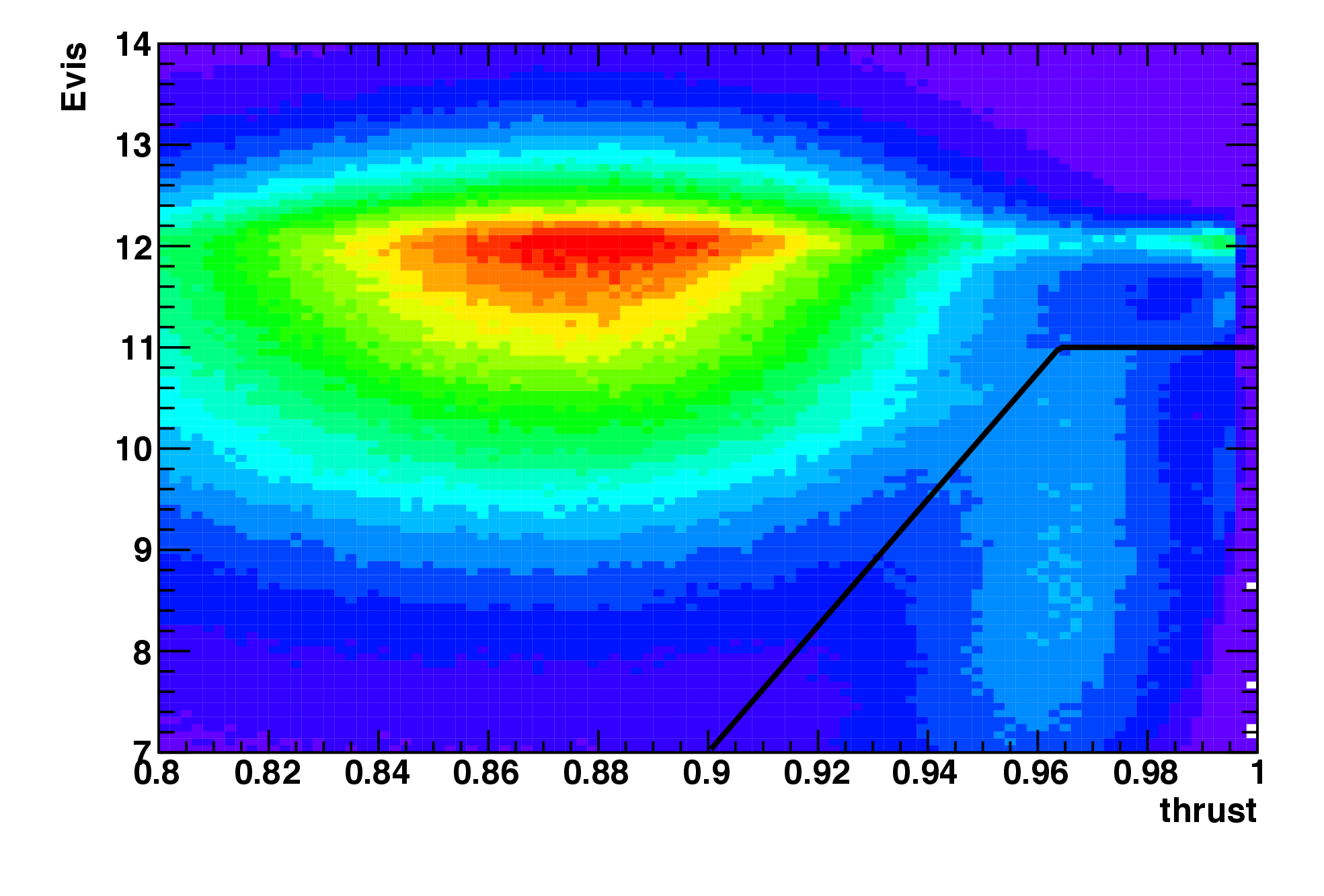}
    {\boldmath
      \put(-38,100){(a)}
    }
    \includegraphics[width=0.43\textwidth]{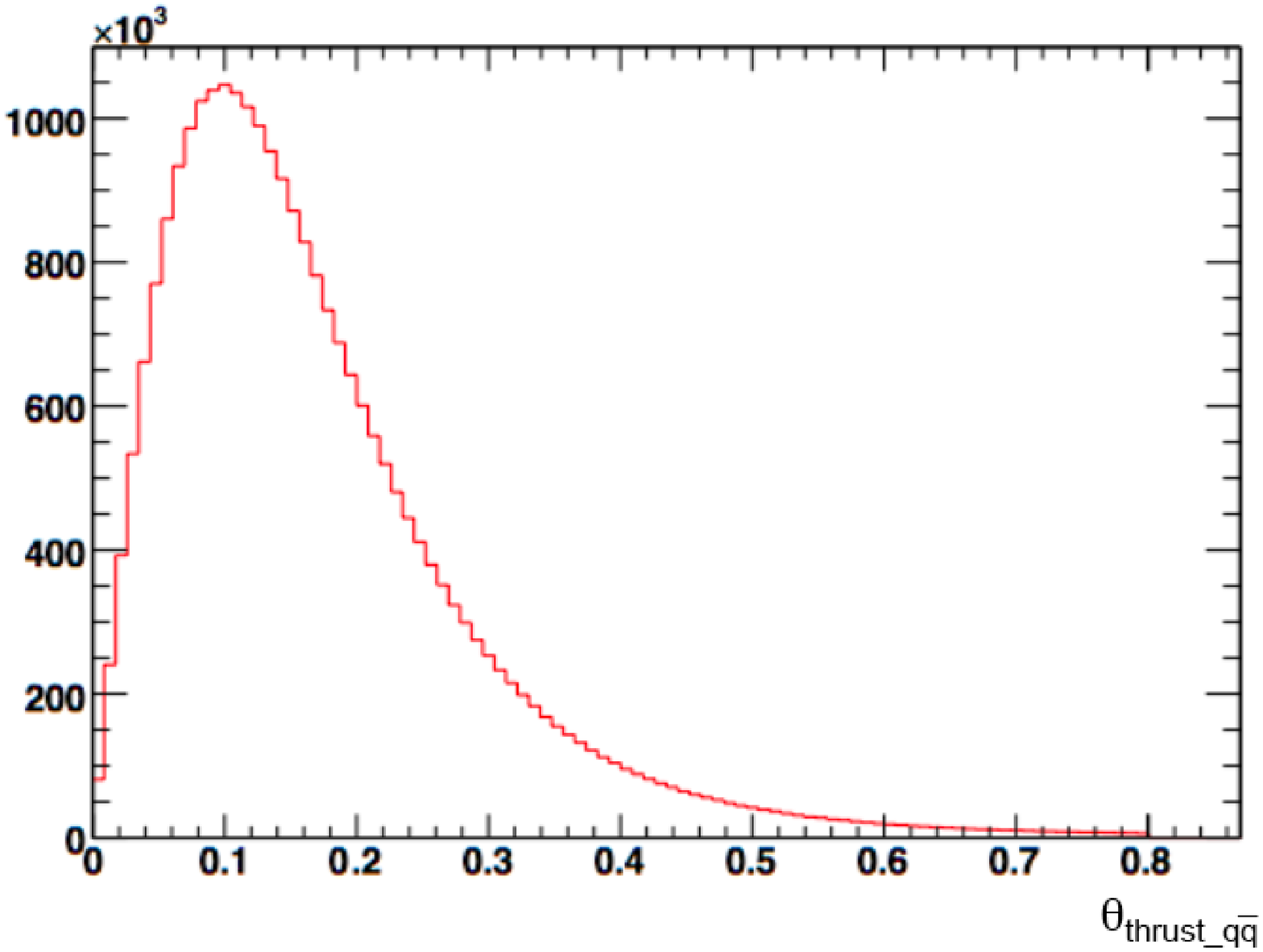}
    {\boldmath
      \put(-38,100){(b)}
    }
      \caption{
      (a) Total visible energy of the events versus the thrust value.
      The peak at about 12 GeV of energy and very high thrust values is due 
      to BhaBha and $\mu^+\mu^- (\gamma)$ events, while the small accumulation visible
      at lower energies and thrust higher than 0.94 is due to $\tau^+\tau^-$ events.
      (b) Opening angle between the thrust axis and the $q\overline{q}$ axis.
    }  
    \label{fig:thrustEvis}
  \end{center}
\end{figure}

\begin{itemize}
\item[-] $e^+e^- \rightarrow \tau^+\tau^-$ contribution. 
Since weak decays are well described in Monte Carlo, 
any azimuthal asymmetries should be visible in simulated events.
The study of simulated $e^+e^- \rightarrow \tau^+\tau^-$ gives an asymmetry consistent 
with zero.
In addition we consider a $\tau$-enhanced data sample, sitting 
in the lower-right side of the $E_{tot}$ vs $T$ distribution of fig.\ref{fig:thrustEvis}(a).
Taking into account the contamination of 25\% from $e^+e^-\rightarrow q\overline{q}$
in this enhanced sample,
the fitted asymmetries are consistent with zero in both reference frames.
The measured asymmetry is then corrected following the equation
 $A_\alpha=A_\alpha^{measured}/(1-D_\tau)$,
 with the fraction of pion pairs ($D_\tau$) from $\tau^+\tau^-$ events 
 estimated using MC to vary 
 from about 1 to 18\% in the individual bins of fractional energy.

\item[-] $e^+e^- \rightarrow c\overline{c}$ contribution.
The fraction of pion pairs due to the $c\overline{c}$ events is much larger
than the $\tau$ component.
Furthermore, both fragmentation processes and weak decays can introduce azimuthal asymmetries.
For this reason, in addition to the $c\overline{c}$ MC sample, we select
a charm-enhanced data sample requiring at least one $D^*$ candidate
from the decay $D^{*\pm}\rightarrow D^0\pi^\pm$, with the $D^0$ candidate
reconstructed only in the decay channel 
$D^0\rightarrow K^-\pi^+$.
We estimate the $c\overline{c}$ contribution in this enhanced sample ($d$) and 
in the main data sample ($D_{charm}$), and we measure the asymmetry in both
samples.
In this way, we are able to extract the Collins asymmetry ($A_\alpha$) 
and the charm asymmetry ($A^{charm}$) solving
the following system of equations:
\begin{eqnarray} \nonumber
A^{meas}_{\alpha} &=& (1-D_{charm})\cdot A_\alpha+D_{charm}\cdot A_\alpha^{charm},\\ \nonumber
A^{D*}_\alpha &=& d\cdot A^{charm}_\alpha+(1-d)\cdot A_\alpha. \nonumber
\end{eqnarray}
 \end{itemize}

\subsection*{Asymmetry dilution due to reconstruction of the thrust axis}
The deviation of the thrust axis from the real $q\overline{q}$ direction 
as shown in fig.\ref{fig:thrustEvis}(b) can lead to a dilution of the measured asymmetries.
This effect could be evaluated using a MC sample. However, the Collins fragmentation functions are not defined in our MC generator. Therefore, we simulate the asymmetries applying different weights to the angular distribution of generated tracks. 
We finally fit the reconstructed azimuthal distributions and we compare the resulting asymmetries with the weights introduced in the simulation.
We find  that, in RF12, the asymmetry is significantly underestimated, and we  apply a correction factor according to the measured value.

\section{Preliminary results}
\begin{figure}[htb!]
\begin{center}
 \includegraphics[width=0.48\textwidth]{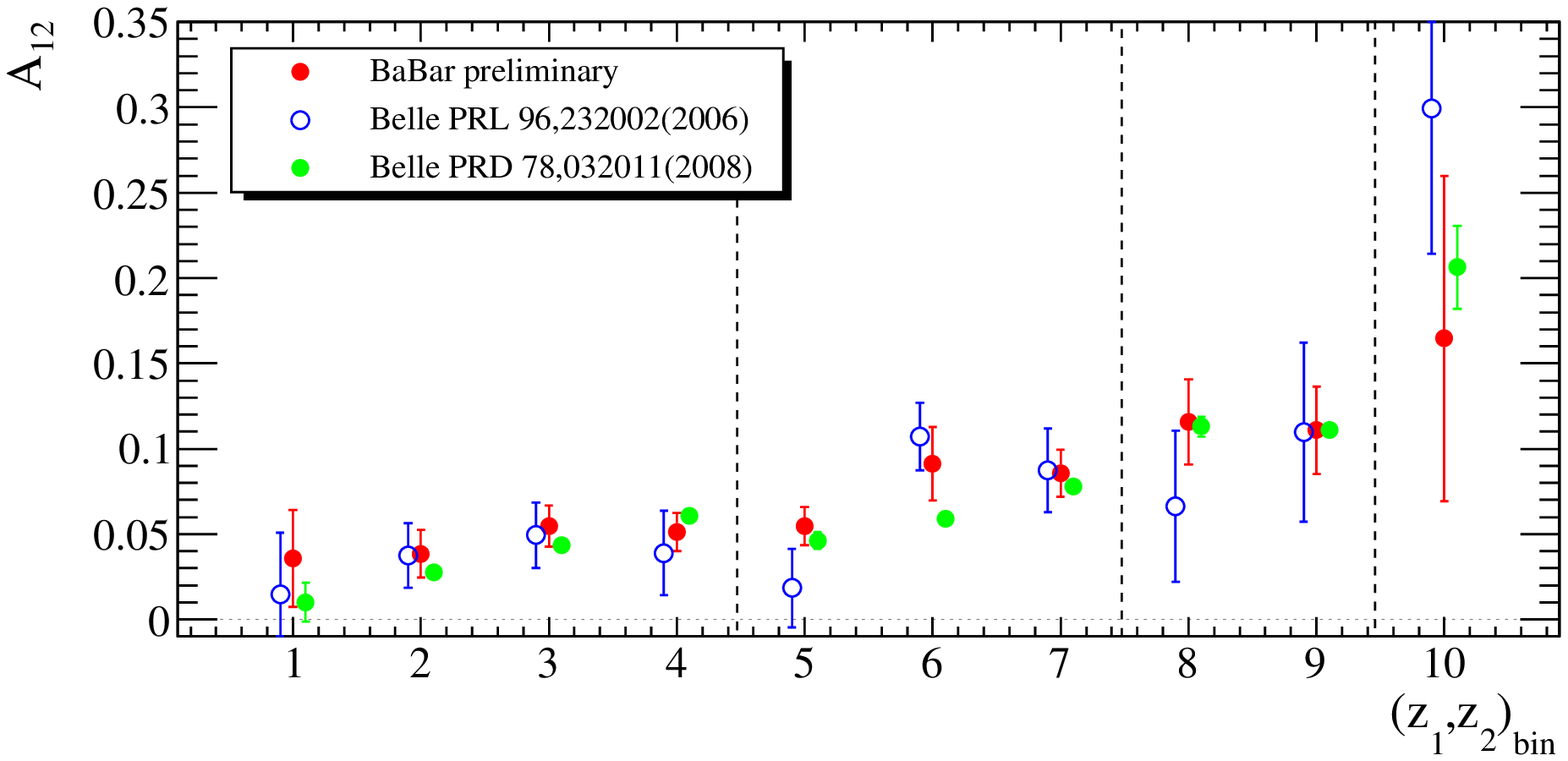}
  {\boldmath
      \put(-150,52){\babar\ Preliminary}
    }
 \includegraphics[width=0.48\textwidth]{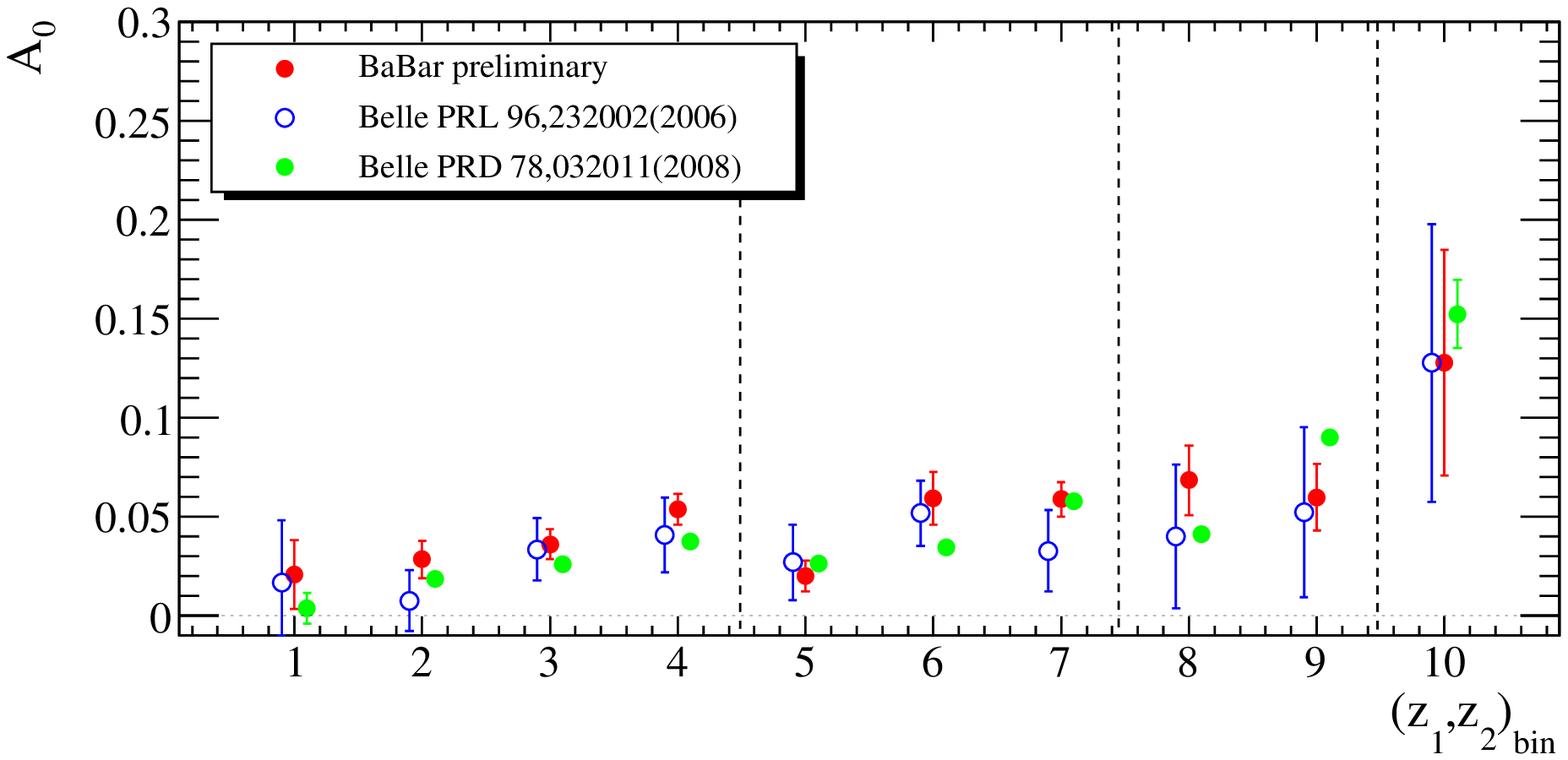}
  {\boldmath
      \put(-150,52){\babar\ Preliminary}
    }
 \caption{Preliminary \babar\ measurement of Collins asymmetries (full circle in red). 
   By comparison the  superseded Belle off-peak results (open circle
   in blue), and Belle results on the full data sample (full green
   circles) are shown. Systematic and statistical errors are
   added in quadrature.} 
 \label{fig:finalOff}
\end{center}
\end{figure}

The preliminary results are presented in 10 symmetric combinations of
($z_1,z_2$) intervals, as described in tab.\ref{tab:zbin}.
All significant systematic errors are evaluated and added in quadrature
for each bins.
\begin{table}[htb!]
\begin{center}
\resizebox{0.8\textwidth}{!}{%
\begin{tabular}{c| c c c c c c c c c c}
$z_2$ &  & & &  & & & & & & \\
     1        &  & & &  & & & & & & \\ 
                &  & 4 & & 7 & & 9 & & \multicolumn{1}{ >{\columncolor{yellow} }r}{10} & &\\ 
     0.7     &  & & &  & & & & & & \\ 
                &  & 3 & & 6 & & \multicolumn{1}{ >{\columncolor{yellow}}r}{8} & & 9 & &\\ 	
      0.5    &  & & &  & & & & & & \\ 
                &  & 2 & & \multicolumn{1}{ >{\columncolor{yellow}}r}{5} & & 6 & & 7 & &\\ 
      0.3    &  & & &  & & & & & & \\ 
                &  & \multicolumn{1}{ >{\columncolor{yellow}}r}{1} & & 2 & & 3 & & 4 & &\\ 
      0.2    &  & & &  & & & & & & \\ \hline
     & 0.2 & & 0.3 & & 0.5 & & 0.7 & & 1 & $z_1$\\
\end{tabular}
}
\caption{Symmetrised $z$-bin subdivision:  we
  divided the asymmetry results in 10 ``bins'' of $z_1$ and $z_2$ as
  summarized above.}
\label{tab:zbin}
\end{center}
\end{table}
In fig.\ref{fig:finalOff} 
 the \babar\ preliminary results
 are compared with the superseded off-peak Belle data
and the off- and on- peak combined Belle data~\cite{ref:belle1,ref:belle}.
We should note that in the newer Belle publication~\cite{ref:belle},
 they estimated a new correction factor due to the approximation
 of the $q\overline{q}$ axis with the thrust axis, so that we correct
 the first Belle results by the
 factor 1.66/1.21.

\begin{figure}[!htb]
  \begin{center}
    \includegraphics[width=0.48\textwidth]{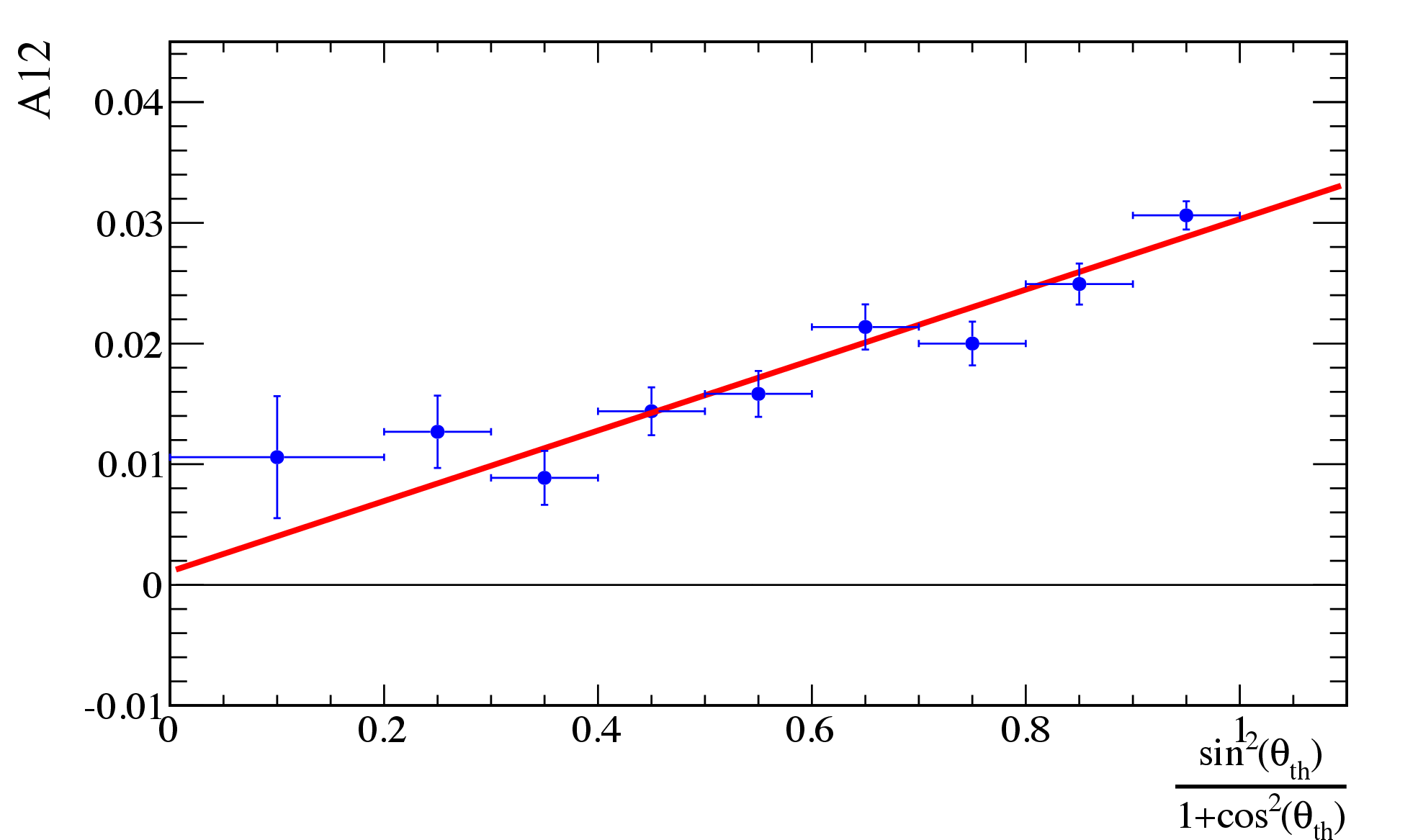}
    {\boldmath
      \put(-30,90){(a)}
    }
    {\boldmath
      \put(-155,90){\babar\ Preliminary}
    }
    \includegraphics[width=0.48\textwidth]{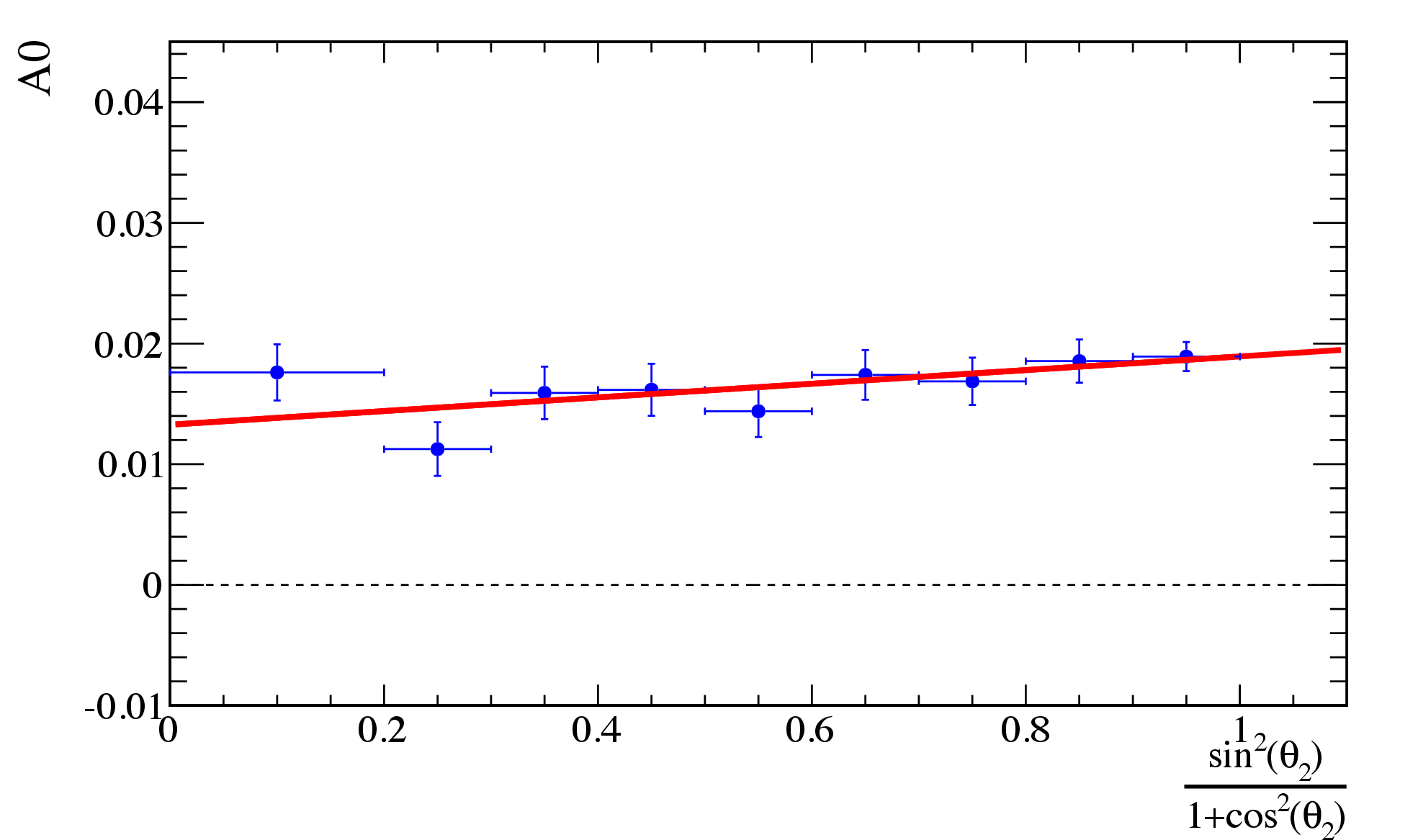}
   {\boldmath
      \put(-30,90){(b)}
    }
     {\boldmath
      \put(-155,90){\babar\ Preliminary}
    }
    \caption{ Collins asymmetry  $A_{12}$ (a), and $A_0$
      (b), as a function of $(\sin^2\theta)/(1+\cos^2\theta)$, where
      $\theta=\theta_T$ and $\theta=\theta_2$ have been used
      in plot (a) and (b), respectively.  }  
    \label{fig:sin2theta}
  \end{center}
\end{figure}  

Following eq.~(\ref{eqn:dr}), we studied also the $\sin^2\theta/(1+\cos^2\theta)$ 
dependence of the Collins
asymmetries in both  reference frames, as reported in fig.\ref{fig:sin2theta}.
The expected $\sin^2\theta_T/(1+\cos^2\theta_T)$ linear dependence 
of asymmetry is observed in the thrust reference frame,
but it seems not holding in the second hadron momentum frame,
where $\theta=\theta_2$ is taken.
The same dependences are observed by Belle in ref.~\cite{ref:belle},
and explained by the fact that the thrust axis describes the original
$q\overline{q}$ direction better than the second hadron momentum.

\section{Conclusions} 
We reported the  preliminary \babar\ results of  Collins asymmetries
in the pion system,
performed using a data sample of about 45 fb$^{-1}$ collected
at the energy of 10.54 GeV.
The asymmetries are studied in function of symmetric bins ($z_1,z_2$)
of the pion fractional energies and in function of 
$\sin^2\theta/(1+\cos^2\theta)$, and are compared with the Belle 
analysis.
The results are in overall good agreement each other.
However, the off-peak data sample is statistically limited, and the update of
the measurement with the full \babar\ data sample is ongoing.


\end{document}